\documentclass[sigconf]{acmart}
\usepackage{booktabs} 
\usepackage{pgfplots} 
\usepackage{balance}

\acmConference[The Web Conference]{The Web Conference}{April 2018}{Lyon, France}
\copyrightyear{2018}
\acmYear{2018} 
\setcopyright{iw3c2w3}
\acmConference[WWW 2018]{The 2018 Web Conference}{April 23--27, 2018}{Lyon, France}
\acmBooktitle{WWW 2018: The 2018 Web Conference, April 23--27, 2018, Lyon, France}
\acmPrice{}
\acmDOI{10.1145/3178876.3186087}
\acmISBN{978-1-4503-5639-8/18/04}

\begin{document}
\title[Automated Approach to Auditing Disclosure of Third-Party Data Collection]{An Automated Approach to Auditing Disclosure of Third-Party Data Collection in Website Privacy Policies}  

\author{Timothy Libert}
\affiliation{%
  \institution{Reuters Institute for the Study of Journalism}
  \institution{Department of Computer Science}
  \institution{University of Oxford}
  \streetaddress{13 Norham Gardens}
  \city{Oxford} 
  \state{United Kingdom} 
}
\email{public@timlibert.me}

\renewcommand{\shortauthors}{T. Libert}

\begin{abstract}

A dominant regulatory model for web privacy is ``notice and choice''. In this model, users are notified of data collection and provided with options to control it. To examine the efficacy of this approach, this study presents the first large-scale audit of disclosure of third-party data collection in website privacy policies. Data flows on one million websites are analyzed and over 200,000 websites' privacy policies are audited to determine if users are notified of the names of the companies which collect their data. Policies from 25 prominent third-party data collectors are also examined to provide deeper insights into the totality of the policy environment. Policies are additionally audited to determine if the choice expressed by the ``Do Not Track'' browser setting is respected. 

Third-party data collection is wide-spread, but fewer than 15\% of attributed data flows are disclosed. The third-parties most likely to be disclosed are those with consumer services users may be aware of, those without consumer services are less likely to be mentioned. Policies are difficult to understand and the average time requirement to read both a given site's policy and the associated third-party policies exceeds 84 minutes. Only 7\% of first-party site policies mention the Do Not Track signal, and the majority of such mentions are to specify that the signal is ignored. Among third-party policies examined, none offer unqualified support for the Do Not Track signal. Findings indicate that current implementations of ``notice and choice'' fail to provide notice or respect choice.

\end{abstract}

%
%

\ccsdesc[500]{Security and privacy~Human and societal aspects of security and privacy}
\ccsdesc[500]{Security and privacy~Usability in security and privacy}

\keywords{Web Privacy, Web Security, Internet Policy, Internet Regulation}

\maketitle

\section{Introduction}

	Although many users may not be aware, web pages are not unitary objects downloaded directly from the party listed in a browser's address bar.  Rather, most web pages are a collection of media elements which are either downloaded from the first-party a user is aware of, or from third-parties a user may not know of.  When a page includes third-party content, HTTP ``Referer'' headers convey the address of the page the user is currently visiting to third-parties.  While users may be happy to see third-party content in a web page, they may not be happy that such content allows third-parties to create records of their browsing behaviors.  The process of using third-party HTTP Referer headers to observe users' web browsing is often referred to as ``web tracking''.

	Research has demonstrated that web pages often expose a user's browsing history to numerous third-parties in tandem \cite{krishnamurthy-2006-generating,libert-2015-1m-tracking,englehardt-2016-million_track,lerner-2016-internet}.  Quite often these parties collect data on users' behavior so that they may be shown advertisements which are tailored to their interests.  The benefit of this system is that users may enjoy learning about products and services relevant to their lives, website operators are able to make the most efficient use of limited screen space, and vendors are able to directly reach potential customers.  However, these commercial imperatives come at a cost to personal privacy and are poorly regulated.

	Despite the fact that half of the top ten companies in the world ranked by market value conduct web tracking, there is very little formal oversight of the practice.\footnote{According to Statistica, Alphabet (Google), Microsoft, Amazon, Facebook, and Tencent are all in the top ten: https://www.statista.com/statistics/263264/top-companies-in-the-world-by-market-value/}  One reason for this is that the United States lacks a top-level data protection authority.\footnote{Although the U.S. Federal Trade Commission has been involved in online privacy for years, the primary remit of the agency is not data protection.}  While nations in the European Union have designated data protection authorities, researchers have found that third-parties routinely ignore regulations designed to police the use of third-party tracking cookies and deem the approach a ``failure'' \cite{trevisan-2017-uncovering}.  The data collection industry claims that formal oversight is not needed due to the adherence to a ``self-regulatory'' framework called ``notice and choice''.
	
	Under the notice and choice framework, users are theoretically notified that data collection is taking place and given options to control the practice (often called an ``opt-out'').  According to industry group Network Advertising Initiative, member companies follow a code which ``requires notice and choice with respect to Interest-Based Advertising, limits the types of data that member companies can use for advertising purposes, and imposes a host of substantive restrictions on member companies' collection, use, and transfer of data used for Interest-Based Advertising'' \cite{nai-2013-conduct}.  Despite the fact that targeted advertising now supports the vast majority of online publications, social media sites, and search engines, there has been very little auditing to determine if self-regulatory frameworks are in fact being followed.  A 2011 evaluation of compliance with industry-defined guidelines for notice and choice conducted by Komanduri et al is one of the most notable studies of the topic \cite{komanduri-2011-adchoices}.

	Although a large volume of academic literature has identified the parties which collect data on websites and deficiencies in the nature of online privacy policies, there has been virtually no attempt to determine if the parties that collect data on a given site are disclosed in the policy of that site.  This study presents the first attempt at auditing disclosure of third-party data flows in website privacy policies and a new software tool, {\tt policyxray}, is presented.  {\tt policyxray} facilitates large-scale auditing of privacy policies and has been used to determine if policies for 207,000 websites accurately disclose the third-parties which collect user data.  
	
	Privacy policies are also analyzed to determine if the text is easy to understand, how long the text would take to read, and if the ``Do Not Track'' choice mechanism is respected.  Network traffic is inspected to determine if transport encryption is used.  Finally, rather than treating third-parties as an undifferentiated whole, the policies and practices of 25 prominent data collectors are examined in order to reveal variations in practices.

	Third-party data collection is wide-spread, but fewer than 15\% of attributed data flows are disclosed.  The third-parties most likely to be disclosed are those with consumer services users may be aware of, those without consumer services are mentioned in less than 1\% of instances.  Policies are difficult to understand and the average time requirement to read both a given site's policy and the associated third-party policies exceeds 84 minutes.   Only 7\% of first-party site policies mention the Do Not Track signal, and the majority of such mentions are to specify that the signal is ignored.  Among third-party policies examined, none offer unqualified support for the Do Not Track signal.  Findings indicate that current implementations of notice and choice fail to provide notice or respect choice.

\section{Research Questions: Notice, Choice, and Security}

	The overarching purpose of this study is to evaluate the efficacy of the notice and choice policy regime.  However, there are no commonly agreed upon definitions of what constitutes sufficient notice or choice on the web.  In the United States this may be partially attributed to the fact that the Federal Trade Commission's guidance on the topic has been ``consistently inconsistent'' \cite{gellman-2016-fair}.  In the European Union, the ePrivacy \emph{Directive} (sometimes referred to as the ``cookie law'') is being replaced by the ePrivacy \emph{Regulation} in 2018 and there remains substantial uncertainty as to what changes will arise from the transition.  
	
	One potential metric for notice is the industry-favored approach of the ``AdChoices'' icon, a small blue arrow which sits in the corner of advertisements.  When clicked upon, the icon will take a user to information about the party responsible for the ad.  However, not all third-parties show ads, thus AdChoices cannot offer full disclosure of all parties collecting data.  Furthermore, researchers have found that ``the purpose of these icons, to provide information to consumers, eluded participants, even when the icons were shown in context on an advertisement'' \cite{ur-2012-smart}.

	It is possible that mentions of sharing data with undefined ``third-parties'' may be viewed as providing notice in the context of a privacy policy.  However, given that users are subject to \emph{both} first- and third-party privacy policies, this type of \emph{ambiguous notice} does not provide users with a means to evaluate all policies to which they are subject.  Indeed, different parties have different policies, and users must know the names of specific data collectors to exercise meaningful choice.  Therefore, for the purposes of this study, merely mentioning ambiguous ``third-party'' data sharing does not qualify as meaningful notice.

	In absence of agreed upon guidelines, purposefully limited questions have been asked in order to determine the degree to which users receive notice and are able to convey choices.  Three decisions have been made to simplify the scope of the evaluation.  First, in place of evaluating inconsistent icons and modal dialogues, this study evaluates human-readable privacy policies as the vehicle for notice.  Second, in order to establish a benchmark for choice, mention of, and respect for, the ``Do Not Track'' (DNT) browser signal is evaluated.  While the online advertising industry has advocated for many different forms of choice ranging from setting ``opt-out'' cookies to instructing users to disable third-party cookies,\footnote{There is significant variability in how users may ``opt-out'' of tracking and opting-out of one service may mean opting-in to another.  For example, Criteo is one of many services which require setting an opt-out cookie, and state in their policy that: \emph{``if your browser settings prevent the use of...third party cookies in general, the choice mechanisms offered by the platforms above will not operate properly.''}  Conversely, Oracle instructs users to turn off third-party cookies to opt-out of tracking, which would have the effect of disabling the Criteo opt-out: \emph{``Oracle does not uniformly process do-not-track signals from browsers. However, you may prevent Oracle from collecting Interest Segments using Cookies on a browser by blocking third-party Cookies in that browser.''}  Thus, Criteo and Oracle have policies which are fundamentally incompatible on a technical level.  However, if both Criteo and Oracle interpreted ``Do Not Track'' as an opt-out, this contradiction could be easily resolved.} DNT is the only signal common to all major browsers and its development was encouraged by the U.S. Federal Trade Commission \cite{ftc-2012-dnt}.\footnote{It is important to note that the F.T.C. also stated that ``work remains to be done on Do Not Track'' in 2012, and such work remains unfinished in 2018.\cite{ftc-2012-dnt}}  Finally, in regards to the security of data transmission, the use of Secure Sockets Layer (SSL) transport encryption is measured.

	Based on the above scoping decisions, the research questions being asked regarding \textbf{Notice} are as follows:
	
	\begin{itemize}
		\item Who are the third-parties which collect user data on websites, and do they have consumer services users may already be aware of?

		\item If users read a privacy policy from a given website will they learn of the specific third-parties which receive their data?
		
		\item How time-consuming and difficult is it to understand website privacy policies?
	
		\item How time-consuming and difficult is it to understand third-party privacy policies?
	\end{itemize} 
	
	Questions regarding \textbf{Choice} are as follows:
	
	\begin{itemize}
		\item Do website privacy policies mention and respect Do Not Track signals?
		
		\item Do third-party privacy policies mention and respect Do Not Track signals?
	\end{itemize}

	Questions regarding \textbf{Security} are as follows:

	\begin{itemize}
		\item What percentage of websites force encrypted connections?

		\item What percentage of third-party requests are encrypted?
	\end{itemize}

\section{Methodology}

	While the above research questions are relatively limited in scope, answering them is a non-trivial task and new methods have been developed for this study.  The first task required is to determine the third-parties which collect data on a given website and if the data transfer is secure.  For this task, the {\tt webxray} software platform is used to monitor third-party network traffic generated by loading a given web page and attribute such traffic to the entities which receive the data.  Second, website privacy polices must be identified, extracted, and audited.  For this task, a new module for {\tt webxray}, named {\tt policyxray}, has been developed.  Although Cranor et al have previously automated analysis of financial website policies \cite{cranor-2016-large}, {\tt policyxray} is the first tool capable of auditing disclosure of specific third-party data flows in website privacy policies and represents a step forward in the automation of privacy policy analysis.  Finally, the relevant policies of third-party data collectors must be selected using a manual process.  These steps are described below.

\subsection{webxray}
	{\tt webxray} is a software platform which measures data leakage to third-parties when loading a given website.  {\tt webxray} leverages an extensive hand-curated library which attributes ownership of third-party domains to the services and corporate entities which control them.  {\tt webxray} has previously been used in academic research \cite{libert-2015-1m-tracking,libert-2015-healthtracking,hauschke-2016-third,yu-2017-jobs} and the {\tt webxray} attribution library has been used to augment findings in other platforms such as OpenWPM \cite{englehardt-2016-million_track,englehardt-2018-email}.
	
	To use {\tt webxray}, one must first generate a list of web pages which are then loaded in a web browser.  During page loading, HTTP element request and receive events are monitored.  To determine privacy leakage, third-party requests are identified by comparing the domain of the page (e.g. ``example.com'') to the domain of the request (e.g. ``tracker.com'').  Sub-domains are ignored so that a request to a domain such as ``images.example.com'' is not recorded as a third-party.  There is no purely automated mechanism to disambiguate between site-specific sub domains and country-specific sub-domains (e.g. ``example.co.uk''), so the Mozilla Public Suffix list is used for this task.\footnote{See https://publicsuffix.org for additional details.}
	
	Once third-party domains are identified, {\tt webxray} searches for them in an internal database of domain ownership.  The {\tt webxray} database is the product of years of detective work as automated tools such as {\tt whois} are unable to reveal the owners of anonymously registered domains.  The process for determining domain ownership is often laborious, but focused human attention produces results not currently achievable by purely machine-driven approaches.\footnote{For example, the owner of one domain was only determined after locating obscure developer documentation.}  The {\tt webxray} attribution database has been modified for this project to reveal the hierarchy of ownership which connects a service to a parent company.  For example, {\tt webxray} is able to determine that the domain ``convertro.com'' is owned by Convertro, which is a subsidiary of Aol, which is a subsidiary of Oath, which in turn is owned by American telecommunications giant Verizon.

	{\tt webxray} currently supports both the Google Chrome browser and the PhantomJS headless browser.  Chrome has the benefit of being the same browser many users employ and is suitable for small volumes of pages.  Due to non-trivial resource requirements and instability when many instances are run in parallel, Chrome is poorly suited for large volumes of pages.\footnote{This is true even when running Chrome in headless mode.}  For this study, the headless browser PhantomJS has been used.  On a suitably robust machine, 64 parallel instances of PhantomJS can be easily run.

	A computer located at an academic institution in the United States is used to conduct measurement.  Using a computer on a university IP block produces better measures than using a cloud hosting provider such as Amazon Web Services due to the fact that IP addresses from cloud hosts are often blocked as they may be used for site scraping and click-fraud.  A major strength of this study is that a cloud service is not used for measurement tasks.  \footnote{For example, the Google Scholar webpage is easily accessible from a university IP, but loading the same page from a cloud service IP address results in a block.}

\subsection{policyxray}

	{\tt policyxray} is a newly developed module for {\tt webxray} which extracts privacy policies and audits their content for disclosure of the specific third-parties which collect data on a given page.  It is the first tool designed to audit observed third-party tracking in website privacy policies and represents the most significant contribution of this study.
		
	{\tt policyxray} relies on a modification to the {\tt webxray} software which facilitates the harvesting of privacy policy links.  When {\tt webxray} loads a page, it extracts all of the links on the page.  The text of each link is evaluated to see if it contains the sub-string ``privacy policy''.  The first such mention is recorded as the policy link and searching stops.  If there is no match, the following strings are searched for in order: ``privacy'', ``terms of service'', ``terms of use''.  Given that policy links are usually found in the footer of a page, links are evaluated in a bottom-to-top order relative to page layout.

	When {\tt policyxray} is run, it attempts to load the URL corresponding to a given site's privacy policy.  Next, {\tt policyxray} attempts to extract the policy text from the page so that it may be evaluated independently of other page elements such as sidebars or footers.  This is necessary because social media companies such as Facebook and Twitter are often mentioned in the text of a footer link, but may not be present in the policy.  Figure 1 illustrates an example of how policy text differs from page text.
	
	\begin{figure}
		\includegraphics[height=3in, width=2.5in]{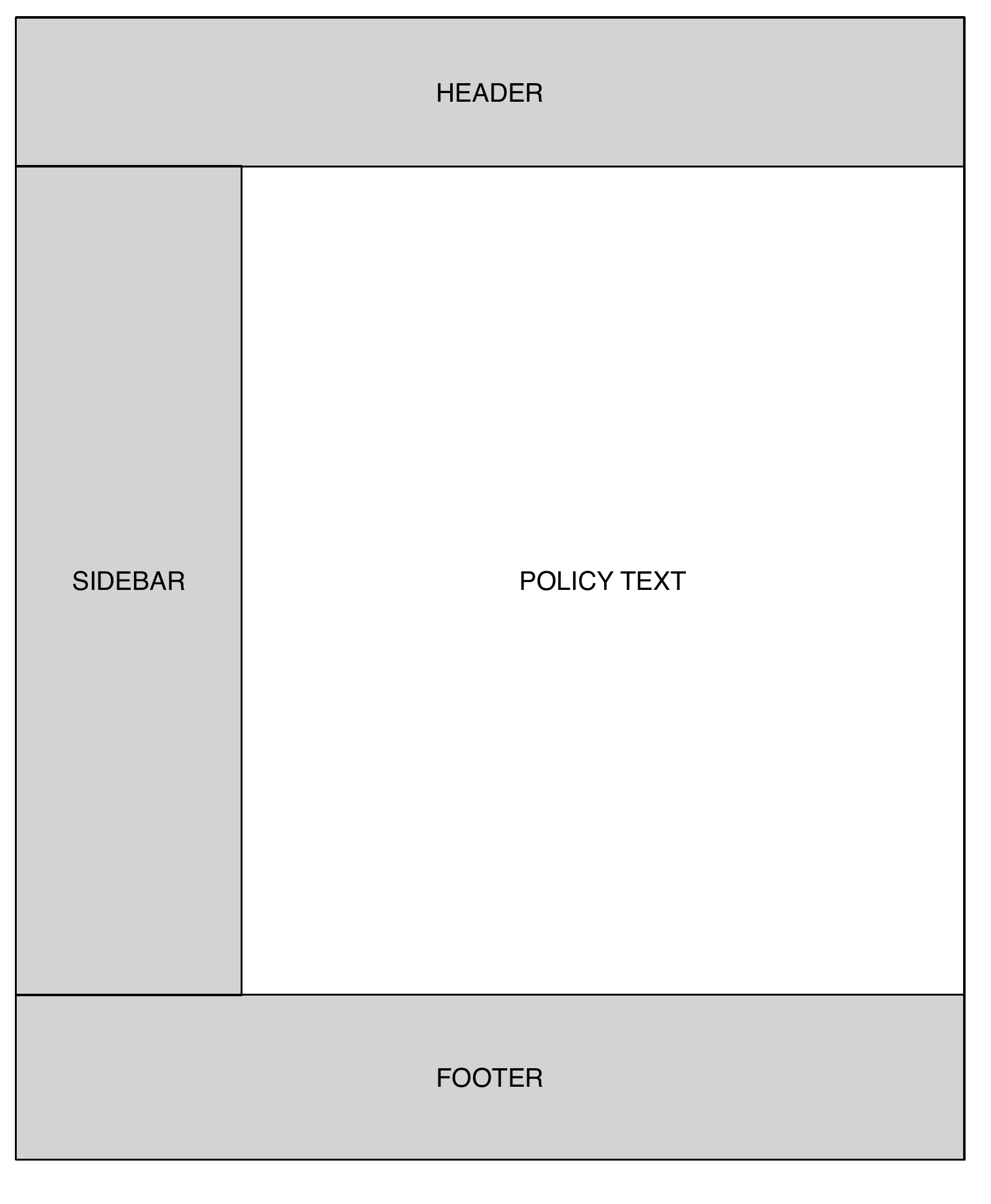}
		\caption{Footer blocks frequently include the names of social media companies.  Such mentions may generate a false positive that the company is mentioned in the policy; therefore {\tt policyxray} extracts only the policy text of a page.}
	\end{figure}
	
	Extracting policy content from a web page is a difficult problem given variations in website coding styles.  To overcome this, the Readability.js Javascript library is used.  Readability.js is an open-source project maintained by Mozilla which provides an automated method for extracting page content by removing boilerplate sections such as page headers and footers.  In order to leverage Readability.js, {\tt policyxray} loads a page with either Chrome or PhantomJS, injects the Readability script into the page, executes it, and strips any remaining HTML elements from the text.
	
	Readability uses a measure of ``link density'' to determine if a given page section is likely to be a navigational block, and therefore not a part of the article (see  \cite{kohlschutter-2010-boilerplate} for an extended discussion of link density).  An earlier incarnation of the Readability.js library was produced by Arc90 Labs; research from 2010 found that version of Readability performed with 95.90\% accuracy \cite{spengler-2010-document}.  Readability is in active development and both Firefox and Safari use it for their ``Reader View'' features.

	To verify that the extracted policy text is in fact a privacy policy, a basic sanity check is done to verify that the page title or text contains the strings ``privacy'' or ``cookie''.  Furthermore, manual inspection of a random sample of harvested text found that 100\% of samples contained \emph{only} policy text (95\% confidence with +/- 5\% interval).

	The main purpose of {\tt policyxray} is to determine if the parties identified collecting data on a given page by {\tt webxray} are disclosed in that page's privacy policy.  To do so, the name of each domain owner is searched for in the policy text, if it is found, it is counted as disclosed.  If a given owner is not found, {\tt policyxray} recursively searches for mention of any parent organizations.  Thus, if ``convertro.com'' is found on a page by {\tt webxray}, {\tt policyxray} will search for the strings ``Convertro'', ``Aol'', ``Oath'', and ``Verizon''.  In cases where there may be variations in the name of a service, such as ``DoubleClick'' and ``Double Click'', both are searched for.  This process allows for a purposefully inclusive approach to auditing disclosure and is designed to give as many opportunities as possible for disclosure to be observed.\footnote{The company ``Inform'' was excluded from analysis due to the fact the word ``inform'' appears with high frequency independently of the company being disclosed.  Including ``Inform'' vastly skews overall findings.}

	To determine if the ``Do Not Track'' (DNT) standard is mentioned and respected in a policy, the string ``do not track'' is searched for in the policy text.  This step is easy to automate, but determining if the string is in reference the DNT standard, and if the choice signal is respected, is a difficult task.  For this study, a random sample of policies with a match on ``do not track'' are manually evaluated to determine if the match is a reference to the DNT standard, and if so, if respect for user choice is clear.

	Finally, policies are evaluated to determine how difficult they are to read and the time needed to read them.  In regards to reading difficulty, the well-established Flesch Reading Ease metric is used.  This metric is a means of measuring the difficulty of reading a text written in the English language.  In regards to time taken to read a policy, this study adopts the approach of McDonald and Cranor who ``assumed an average reading rate of 250 words per minute'' \cite{mcdonald-2008-cost}.

\subsection{Selection of Third-Party Privacy Policies}

	There are three main reasons a third-party may be collecting data on a given website.  First, the party may be a Content Delivery Network (CDN).  In this case, data collection may be viewed as incidental and largely outside the scope of notice and choice.  Second, the party may be a service used for Distributed Denial of Service (DDoS) mitigation and data collection may also be viewed as incidental.  In the final case, a third-party may be collecting user data for audience tracking, online advertising, data brokerage, or other tasks which require processing and storing data related to the behavior of specific users.\footnote{The term ``natural persons'' may also apply here.}
	
	Once {\tt webxray} has produced a ranked report of the third-parties most frequently found in the studied population of sites, those parties which are CDNs and DDoS mitigation services are excluded from further consideration.  For the remaining parties, the most salient privacy policy is chosen manually.  The considerations in this process are first to choose the policy which is most applicable to third-party data collection, is written in English, and if there are policies for more than one country, the U.S. policy is used to reflect the location of the machine being used for the study.  Once the policies are selected, it is possible to isolate the policy text, identify Do Not Track clauses if present, and evaluate readability.

\subsection{Limitations}

	While the methods perform well at scale, they are not without limitation.  First, because PhantomJS was used as a browser, requests for Flash elements may be missed and thus a given company collecting data will not be identified.   Prior research has observed that PhantomJS may not successfully load some pages \cite{englehardt-2016-million_track}, but user-agent randomization greatly reduces this issue.  Likewise, due to rapid ingestion of pages, it is possible that the IP address used for collection be black listed for appearing to be a ``bot'' and pages and elements will not load.  
	
	Second, due to the nature of real-time ad bidding, websites which rely on advertising will likely expose users to different parties on each page load depending on what parties win a given auction.  Thus, for sites with advertising, loading pages a single time will produce an under-count of the number of parties which may collect user data on the site.  However, because {\tt webxray}'s database of domain ownership primarily contains major ad networks rather than small clients, and {\tt policyxray} only searches for identified parties, variability in the long-tail of trackers may not have an outsized effect on overall findings related to disclosure.  Nonetheless, it is important to point out that the number of parties being searched for is \emph{fewer} than the total number of parties present.
	
	Third, although Readability.js is used by major browsers and has been tested in prior research, it is not perfect and it is possible portions of a policy may not be extracted.  If such portions contain the only mention of a given third-party, that will produce a false negative.  Conversely, if extraneous non-policy text is included, and that text includes the a mention of a given third-party, it may produce a false positive.  However, as noted above, a sample of collected policies detected no such issues.
	
	Finally, due to the fast pace of ownership changes in the online advertising market, it is possible that some parties may have new parents or subsidiaries which are not yet reflected in the {\tt webxray} database.  While the above limitations may impact findings, this study nonetheless represents the first attempt to perform the task of auditing the disclosure of third-party data flows in website privacy policies at scale.

\section{Research Findings}

	In October 2017, a computer based at a United States academic institution is used to scan one million popular websites as identified by the Alexa company using the {\tt webxray} software platform.  Of these pages, 938,093 are successfully loaded and privacy policy links for 248,029 pages are extracted.  {\tt policyxray} is used to extract 184,897 unique policies corresponding to 207,000 sites.  The number of policies is lower than the number of sites because sites owned by a single entity often share policies.  Of the most prevalent third-parties receiving user data, 25 are selected for their pertinence to the study of notice and choice and their privacy policies are extracted for analysis.  

	Findings shed light on the general state of tracking on popular websites, the nature of the third-parties collecting user data, rates of disclosure for third-party data flows, the complexity and length of privacy policies, respect for the ``Do Not Track'' (DNT) standard, and the security practices used by popular websites and third-party data collectors.  Taken as a whole, findings demonstrate that there is poor disclosure of third-party data collection, policies are difficult and time consuming to read, DNT is rarely respected, and security practices are suboptimal.  The follow sections address these findings in detail.

\begin{table}
	\caption{Third-Party Prevalence, SSL Use, and First-Party Disclosure\\ 
		\emph{\dag Denotes Company has Consumer Services}
	}
	\label{3ps}
		\begin{tabular}{lllrr}
		\toprule
		Company & \% Tracked & \% SSL & \% Disclosed\\
		\midrule
		Google \dag		& 82.81	& 80.35 & 38.29 \\
		Facebook \dag	& 33.37	& 91.61 & 17.50 \\
		Twitter \dag	& 12.26	& 90.43 & 10.74 \\
		AppNexus		& 11.97	& 59.98 & 0.44 \\
		Oracle			& 11.21	& 41.51 & 3.72 \\
		Adobe \dag		& 10.14	& 70.48 & 5.77 \\
		Oath \dag		& 9.67	& 57.64 & 4.42 \\
		The Trade Desk	& 7.38	& 56.49 & 0.12 \\
		Acxiom			& 7.10	& 34.21 & 0.26 \\
		Rubicon Project	& 6.68	& 71.62 & 0.12 \\
		OpenX			& 5.78	& 52.50 & 0.29 \\
		Lotame			& 5.71	& 29.82 & 0.29 \\
		IPONWEB			& 5.64	& 66.11 & 0.07 \\
		Casale Media	& 5.05	& 63.74 & 0.05 \\
		Criteo			& 4.93	& 62.26  & 2.75 \\
		Neustar			& 4.78	& 40.05 & 0.04 \\
		PubMatic		& 4.61	& 54.27 & 0.19 \\
		Media Math		& 4.60	& 56.23 & 0.04 \\
		Microsoft \dag	& 4.57	& 72.27 & 12.56 \\
		comScore		& 4.57	& 53.42 & 1.74 \\
		Nielsen Online	& 4.03	& 41.41 & 0.35 \\
		AdForm			& 3.96	& 50.71 & 0.88 \\
		New Relic		& 3.94	& 97.18 & 0.60 \\
		Quantcast		& 3.71	& 46.01 & 1.46 \\
		Rocketfuel		& 3.65	& 59.83 & 0.10 \\
		\bottomrule
	\end{tabular}
\end{table}

	\subsection{Current State of Web Privacy}
	
	Prior work has investigated the state of tracking on the Alexa top one million websites in both 2015 and 2016 \cite{libert-2015-1m-tracking,englehardt-2016-million_track}.  It is therefore useful to briefly provide an overview of the current state of web tracking in order to contribute to the historical record.  
	
	Of the 938,093 pages which loaded successfully, 91.27\% initiate a request to download a third-party page element, thereby potentially exposing users to cross-site tracking.  Pages which initiate a third-party request expose users to an average of 10.89 unique domains per page-load.  The number of third-party requests for the top 10,000 sites is 20, whereas for the bottom 10,000 it is 10 (see Figure 3 for detail).  70.60\% of page loads result in the setting of a third-party cookie, and pages with third-party cookies have an average of 11.24 distinct cookies per-page.  86.84\% of pages include Javascript code loaded from a third-party domain.  

	\subsection{Identification of 25 Prominent Third-Party Data Collectors}
	
	As noted above, {\tt webxray} uses a database of domain ownership which provides a hierarchical means to trace ownership of data collected by third-parties on the web.  Table 1 shows 25 of the most prominent data collectors discovered on the Alexa top one million websites.  The parties are chosen because they are primarily active in the processing and storing of the data of users rather than content hosting or DDoS mitigation.  Likewise, all of the parties chosen set third-party cookies which may be used for cross-site tracking.  In cases where nearly all of a single company's data was traced back to a subsidiary, the subsidiary was selected in place of the parent.  This was the case with Google (an Alphabet subsidiary) and Oath (a Verizon subsidiary).

	Table 1 shows the percentage of sites which may be tracked by a given third-party.  It is important to note that that for all of the companies chosen, the percentage of sites tracked is a composite measure.  For example, if Aol and Yahoo are on the same site, the site is counted once each for Aol, Yahoo, and Oath rather than twice for Oath.  These composite measures provide insight into the reach of various companies.  For example, Google tracks over 82\% of sites, Facebook over 33\%, and Twitter over 12\% \footnote{It is interesting to note that Twitter is down from 18\% in May 2014, and this is a trend worth exploring if the social network declines in relevance \cite{libert-2015-1m-tracking}.}.  Table 1 also illustrates a long-tail distribution of third-party data collectors.  The fifth place company, Oracle, tracks 11.21\% of sites, a fraction of Google.  Likewise, the 25th place company, Rocketfuel, tracks 3.65\%, a fraction of Oracle.

	Returning to the evaluation of notice, it may be assumed that if users have pre-existing consumer relationships with a company they may already be familiar with data collection practices.  For example, Twitter's privacy policy states that ``We may personalize the Services for you based on your visits to third-party websites that integrate Twitter content such as embedded timelines or Tweet buttons''.\footnote{https://twitter.com/en/privacy}  Thus, from a perspective of notice and choice, consumer services such as social media, search, and email may theoretically have provided notice of data collection independently of site policies.  
	
	Of the 25 companies examined, only six have consumer services.  Therefore, for the majority of third-party data collectors there is virtually no chance users will have awareness of data collection practices due to prior interaction with a service.  It is also worth noting that just because a company has a consumer service that does mean that all users who may be tracked are users of the service.  For example, people who do not use Twitter, and have no reason to read Twitter's privacy policy, may still be tracked by Twitter.
	
\begin{table}
	\caption{
		Third-Party Privacy Policy Characteristics \\
		\emph{
			\dag Denotes DNT Mention \\ 
			\ddag Denotes DNT Partially Respected
		}
	}
	\label{3ps}
		\begin{tabular}{lllrr}
		\toprule
		Company & Word Count &  Reading Ease \\
		\midrule
		Google				& 2773	& 39.67 \\
		Facebook			& 2701	& 48.94 \\
		Twitter				& 3799	& 35.1  \\
		AppNexus			& 3901	& 43.22	\\
		Oracle \dag			& 4844	& 29.18 \\
		Adobe				& 1700	& 29.08 \\
		Oath \dag			& 2461	& 35.61	\\
		The Trade Desk \dag	& 5731	& 39.06	\\
		Acxiom				& 881	& 26.61 \\
		Rubicon Project		& 720	& 37.84 \\
		OpenX  \dag\ddag	& 3345	& 35.31 \\
		Lotame \dag\ddag	& 3150	& 29.48	\\
		IPONWEB				& 947	& 29.48 \\
		Casale Media		& 1301	& 25.90 \\
		Criteo \dag			& 3287	& 38.25 \\
		Neustar				& 5903	& 31.31 \\
		PubMatic \dag		& 4360	& 18.42 \\
		Media Math			& 4794	& 39.16 \\
		Microsoft \dag		& 25367	& 40.89 \\
		comScore \dag		& 873	& 35.27 \\
		Nielsen Online		& 1566	& 42.41 \\
		AdForm				& 2134	& 26.85 \\
		New Relic			& 4150	& 43.06 \\
		Quantcast			& 2924	& 40.79 \\
		Rocketfuel			& 3445	& 35.10 \\
		\bottomrule
		Average			& 3882 & 35.48 \\
	\end{tabular}
\end{table}

	\subsection{Disclosure of Companies in Privacy Policies}
		
	The crux of this study is an evaluation of whether or not website privacy policies provide notice of the third parties which collect data on a given site.  As detailed in the methods section, {\tt webxray} is used to determine the third-parties which collect user data on a given site.  For 207,000 websites, {\tt policyxray} is used to verify if these companies are mentioned in the site's privacy policy.  A total of 1,807,491 instances of data transmission to a known third-party are audited.  It is found that only 14.80\% of data transmissions to identified third-parties are disclosed.  Users who read website privacy policies are therefore very unlikely to be notified of the parties which collect their data.
	
	While the overall rate of disclosure is low, it is not uniform across parties.  As Table 1 shows, transfers to Google are disclosed in 38.29\% of cases.  While over 60\% of transfers to Google are \emph{not} disclosed, there is clearly a strong possibility users may learn of data transfer either through a site policy, or through Google's own policies.  For companies with consumer services, disclosure is lowest for the Oath group with 4.42\%.  Again, because Aol and Yahoo are Oath subsidiaries it is possible users are notified via consumer policies.  For all companies with consumer services, the average rate of disclosure is 14.88\%.
	
	Among the 25 prominent third-parties inspected, disclosure for non-consumer services is sharply lower.  For the 19 services which most users are likely unaware of, the average rate of disclosure is less than 1\%.  Simply put, if a user does not have a pre-existing consumer relationship with a third-party there is virtually no chance they will learn of these parties by reading privacy policies.
		
	\subsection{Readability of Policies}
	
	Beyond mere mention of the parties collecting user data, privacy policies for websites may detail a host of other issues related to data storage, retention, and use.  However, in order for this information to be useful, it must be understood by most users.  While the degree to which a given text is understandable relies on a host of factors ranging from a given user's literacy to familiarity with the minutia of data protection regulations, it is possible to use the well-established Flesch Reading Ease (FRE) metric to evaluate the difficulty of reading a given text.  
	
	FRE scores range from 0-100, with lower scores indicating the text is more challenging to read.  Given that it is difficult for individuals to consent to contracts which they cannot understand, the U.S. state of Florida requires that insurance policies are written in a way which generates ``a minimum score of 45 on the Flesch reading ease test''.\footnote{Florida Statutes Section 627.4145 - Readable Language In Insurance Policies. (Fla. Stat. \S 627.4145)}  
	
	Analysis of website privacy policies reveals that they have an average FRE score of 39.83.\footnote{This level of difficulty shows little variation relative to the Alexa rank of a given site (see Figure 2).}  Thus, if an average website privacy policy were an insurance policy in the state of Florida, it would not pass basic legal requirements.

	When turning attention to the privacy policies of the 25 prominent data collectors detailed in Table 2, it is found that the average readability score, 35.48, is lower than that for website policies.  Again, if these policies were for insurance in Florida, rather than online privacy, they would not be valid.  
	
	There is significant variation between various third-party policies.  Facebook has the most readable policy, with a score of 48.94; at the bottom end, PubMatic has a score of 18.42, which makes it more difficult to read than an article in the \emph{Harvard Law Review} \cite{mcdonald-2009-comparative}.

	\subsection{Time Required to Read Policies}
	
\begin{figure}
		\caption{Sites ranked higher in Alexa tend to have longer policies (blue), but Flesch Reading Ease (red) shows much less variation.}
\begin{tikzpicture}

	\begin{axis}[
		xlabel={Alexa Rank Averaged per 10,000},
		xmin=0, xmax=100,
		ymin=1000, ymax=2400,
		xtick={0,20,40,60,80,100},
		ytick=,
		legend pos=north west,
		ymajorgrids=true,
		grid style=dashed,
	]
 
\addplot[
    color=blue,
    mark=none,
    ]
    coordinates {
		(0,2303.083176312248)
		(1,1945.1535915259847)
		(2,1759.4966711051932)
		(3,1799.1148291651991)
		(4,1730.513350883791)
		(5,1675.4213178294574)
		(6,1653.54751827626)
		(7,1630.6232272901495)
		(8,1633.5363673303534)
		(9,1566.1078199052133)
		(10,1621.3374799357946)
		(11,1549.7914979757086)
		(12,1536.6363256784969)
		(13,1572.4013997529848)
		(14,1535.088285960379)
		(15,1508.8425460636515)
		(16,1512.587037806398)
		(17,1484.4966015293119)
		(18,1527.6184602296894)
		(19,1560.4904430929626)
		(20,1536.9060283687943)
		(21,1445.2174490699733)
		(22,1467.8709398007795)
		(23,1470.40581655481)
		(24,1480.2885662431943)
		(25,1517.0117260787993)
		(26,1373.533774208786)
		(27,1465.9098436062557)
		(28,1454.5298013245033)
		(29,1468.6048689138577)
		(30,1404.7591647331787)
		(31,1437.2967332123412)
		(32,1436.1255453223462)
		(33,1431.8168812589413)
		(34,1376.022076092062)
		(35,1401.6225259189443)
		(36,1467.5522531160116)
		(37,1468.365292425695)
		(38,1387.4310776942357)
		(39,1411.9828022255942)
		(40,1385.1743614931238)
		(41,1371.1493475108748)
		(42,1376.4193083573487)
		(43,1360.5169277412836)
		(44,1417.3969620253165)
		(45,1388.3576027736503)
		(46,1355.467313585291)
		(47,1421.2059686888454)
		(48,1403.1830033800097)
		(49,1447.8577055637618)
		(50,1348.8318350079323)
		(51,1369.8890010090818)
		(52,1375.8890633176347)
		(53,1437.93493852459)
		(54,1342.8210361067504)
		(55,1357.4161616161616)
		(56,1379.7028254288598)
		(57,1313.8153768335862)
		(58,1384.8780614903596)
		(59,1369.292960662526)
		(60,1400.9886597938143)
		(61,1465.1712222795256)
		(62,1330.1314405888538)
		(63,1314.4274322169058)
		(64,1361.99044078598)
		(65,1354.5100105374079)
		(66,1366.4327390599676)
		(67,1361.4832985386222)
		(68,1303.4916230366491)
		(69,1339.5800984144341)
		(70,1385.9792008757527)
		(71,1390.6212204507972)
		(72,1423.0788013318536)
		(73,1370.6392719249861)
		(74,1349.9561925365063)
		(75,1372.963503649635)
		(76,1359.1453175457482)
		(77,1364.910447761194)
		(78,1411.465526751241)
		(79,1384.6446644664466)
		(80,1365.8266888150608)
		(81,1350.0870786516855)
		(82,1371.7059161401494)
		(83,1344.2705555555556)
		(84,1349.2939150401837)
		(85,1359.2114188807236)
		(86,1338.8258903335218)
		(87,1383.381425233645)
		(88,1395.6296743063933)
		(89,1369.4768490679494)
		(90,1331.1807526218383)
		(91,1363.8444165621079)
		(92,1363.52131147541)
		(93,1343.9746575342465)
		(94,1318.7557195571956)
		(95,1398.0797592174567)
		(96,1397.9083402146987)
		(97,1410.067978533095)
		(98,1382.975845410628)
		(99,1329.7633069082672)
    };
    \legend{Word Count, Reading Ease (Scaled x30)}

 \addplot[
    color=red,
    mark=none,
    ]
    coordinates {
		(0,1147.549778761062)
		(1,1169.5511730703843)
		(2,1180.037555479686)
		(3,1172.404332129964)
		(4,1154.4414274631497)
		(5,1171.8562874251497)
		(6,1166.1205870686235)
		(7,1172.3914759273875)
		(8,1159.9584199584199)
		(9,1190.3441295546559)
		(10,1178.2367447595561)
		(11,1160.6733167082295)
		(12,1164.61970613656)
		(13,1181.2216624685138)
		(14,1163.0738611233967)
		(15,1168.8134135855546)
		(16,1158.932536293766)
		(17,1141.8676019289785)
		(18,1171.871727748691)
		(19,1212.0322291853176)
		(20,1190.4049135577798)
		(21,1174.863387978142)
		(22,1226.4816472694718)
		(23,1138.443526170799)
		(24,1178.339552238806)
		(25,1177.4419729206963)
		(26,1172.347440944882)
		(27,1183.7357954545455)
		(28,1193.5087719298247)
		(29,1162.3878437047756)
		(30,1161.7097701149423)
		(31,1166.146660439047)
		(32,1192.5641025641025)
		(33,1171.92023633678)
		(34,1178.8235294117649)
		(35,1169.8834385624089)
		(36,1165.4694485842026)
		(37,1181.4776119402984)
		(38,1165.2400835073067)
		(39,1174.6441558441556)
		(40,1172.1892299949673)
		(41,1183.4715284715285)
		(42,1179.3323442136498)
		(43,1169.9684044233807)
		(44,1176.3584415584414)
		(45,1178.9600409836066)
		(46,1167.3728369166229)
		(47,1093.7963891675026)
		(48,1178.4029850746267)
		(49,1166.2290076335878)
		(50,1161.0442864953527)
		(51,1163.6990595611285)
		(52,1148.7216828478965)
		(53,1182.1810481736368)
		(54,1153.6305048335123)
		(55,1176.7436974789914)
		(56,1172.6732673267327)
		(57,1165.8040859088528)
		(58,1166.8843683083512)
		(59,1164.6035125066526)
		(60,1190.6449893390193)
		(61,1185.7468220338983)
		(62,1195.6816947311243)
		(63,1173.8097855964816)
		(64,1124.3006034009875)
		(65,1185.089141004862)
		(66,1187.6302521008404)
		(67,1164.038982133189)
		(68,1184.5751633986927)
		(69,1159.0289608177172)
		(70,1177.661748013621)
		(71,1139.9829642248721)
		(72,1160.0747986191025)
		(73,1075.6260720411665)
		(74,1176.5735047512576)
		(75,1154.7997678467789)
		(76,1169.584026622296)
		(77,1205.7260726072607)
		(78,1145.0682593856654)
		(79,1151.5226101888952)
		(80,1135.6189389617798)
		(81,1147.2902097902097)
		(82,1125.5913978494623)
		(83,1130.7003444316879)
		(84,1160.715976331361)
		(85,1102.342289719626)
		(86,1138.3440986494422)
		(87,1163.6851628468035)
		(88,1164.881693648817)
		(89,1174.9844430616056)
		(90,1155.0095969289828)
		(91,1164.4680851063831)
		(92,1149.8577235772357)
		(93,1193.3262260127933)
		(94,1176.1941448382127)
		(95,1159.6359411309063)
		(96,1178.6479591836735)
		(97,1160.4255319148936)
		(98,1173.8176352705411)
		(99,1136.1761297798378)
    };
\end{axis}
\end{tikzpicture}
\end{figure}
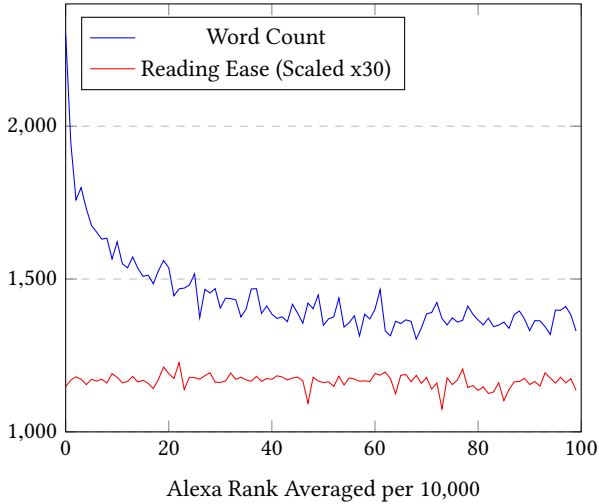

	In 2008, McDonald and Cranor posed an intriguing question: ``If website users were to read the privacy policy for each site they visit just once a year, what would their time be worth?'' \cite{mcdonald-2008-cost}.  By analyzing the ``word count of the 75 most popular websites'', they determined that the ``national opportunity cost'' of reading privacy policies would be \$781 billion dollars.  The present study updates McDonald and Cranor`s findings, vastly expands the number of policies studied, and introduces a new composite measure to determine the time taken to read \emph{both} first- and third-party policies.

	Across policies for 207,000 sites, the average number of words per-policy is 1,404.  Using ``an average reading rate of 250 words per minute'', an average website policy would require 5.6 minutes to read \cite{mcdonald-2008-cost}.  This is lower than the average of 10 minutes found by McDonald and Cranor in 2008.  This is due to the fact that policies for sites ranked higher by Alexa tend to have longer policies, whereas the majority tend to be significantly shorter (see Figure 2).

	One possible reason policies for highly-ranked sites are longer may be that they are written by teams of lawyers who create detailed, custom policies.  In comparison, low-ranked sites likely do not have the resources needed to generate complex policies.  Nevertheless, it appears that when looking at the larger population of sites, the time requirements of reading an average website policy could be much lower than earlier work estimated.  However, this may not be the case.

	As the findings presented thus far make clear, a user visiting a given website is subject to \emph{many} policies: those of the website as well as the third-party data collectors.  As Table 2 details, the average length of a third-party privacy policy is 3,882 words and would require 15.5 minutes to read, nearly three times a website privacy policy.
	
	Because a single site may expose users to several policies in tandem, it is possible to calculate the total time needed to read \emph{all} applicable policies of a given site.  Considering that the total minutes to read applicable site policies is the sum of the word count of the first-party policy ($WC_{fp}$) and the word count of all third-party policies ($WC_{tp}$), divided by 250, the following formula may be derived:
	
	$$t_{mins} = \frac{WC_{fp} + \sum WC_{tp}}{250}$$
	
	Applying this formula reveals that the total time needed to read all applicable policies for a given site is 84.7 minutes on average.  This calculation does \emph{not} take into account that users may not need to re-read a third-party policy on every site they view, and is only applicable to the first site in which the total set of policies is encountered.  
	
	As previous sections detailed, there is a low probability users will be aware of third-party policies to start with, and it is highly unlikely any user would have the ability to locate relevant policies, let alone the time to read them.  Thus, the assertion is not that that users actually spend over an hour reading policies, rather this finding underscores that the notice and choice regime is fundamentally untenable when the full range of policies is considered.

	\subsection{Respect for User Choices}
	
	As the above findings have made clear, the likelihood of a user receiving notice of the third-parties receiving their data by reading a site's privacy policy is remarkably low.  However, it is possible that users could express their choices to control data collection in a way which would not require user notification.  The ``Do Not Track'' (DNT) mechanism accomplishes this task and is a means for users to communicate their desire not to be tracked to parties receiving HTTP requests.  
	
	DNT is a setting available in all major web browsers and is easy for users to enable.  The U.S. Federal Trade Commission has been highly supportive of the standard and encouraged its development \cite{ftc-2012-dnt}.  According to the technical specification, DNT provides a ``means of allowing users to express their preferences about tracking, including to opt out of tracking some or all of the time'' \cite{mayer-2011-DNT}.  
	
	DNT may be viewed as a polite request and there is no technical mechanism to force compliance on the part of data collectors.  Rather, data collectors must commit to respect the expressed choice signal in their policy documents.  This study is the first to examine support of DNT for both websites and third-party data collectors at scale.
	
%
%

	Across the population of website policies analyzed,  8\% contain the string ``do not track''.  A manual analysis of a sample of policies determines if the string is in reference to DNT, and if so, if DNT is honored.  It is found that 15\% of instances of the ``do not track'' phrase are not in reference to DNT (e.g. ``we do not track users'' was a common phrase that is not DNT-related per se).   Thus, when rounding to the nearest integer, 7\% of all policies discuss DNT. 
	
	Of policies mentioning DNT, 77\% explicitly do \emph{not} honor it.  For example, one representative policy stated that ``we will not disable tracking technology that may be active on the Sites in response to any `do not track' requests that we receive from your browser''.\footnote{http://www.cmaworld.com/privacy/}  A policy for a government website in Arkansas specifies that ``While the United States Federal Trade Commission has endorsed DNT, our Sites do not currently support DNT codes.'' \footnote{http://www.arkansas.gov/policies/privacy-policy}  
	
	Only 23\% of policies mentioning DNT contain a clear commitment to honor a user's DNT preference.  One such commitment is found in the following statement: ``We honor do not track signals and do not track, plant cookies, or use advertising when a Do Not Track (DNT) browser mechanism is in place.''  However, it is important to reiterate that such commitments are voluntary and difficult to audit.

	Given that a single website may not have the ability to ``track'' users between sites, the language of DNT may not be fully applicable.  However, for the third-parties which track users across the web, DNT has particular salience.  Furthermore, whereas small website operators may be ignorant of the DNT standard, major third-party data collectors are well aware of it and employ lawyers with expertise in data protection regulations.  
	
	Despite this awareness, only nine of 25 data collectors mention the DNT standard in their privacy policies.  As with first-party disclosures, the majority of these mentions are to specify that DNT is ignored.  Only two third-parties, OpenX and Lotame, offer qualified support for DNT.  Lotame's policy represents the best respect for the spirit of DNT: ``If Lotame receives a `Do Not Track' signal from any browser other than Internet Explorer, Lotame will implement an opt-out.''  Twitter was previously the largest party to respect DNT, but has recently stopped doing so.

	\subsection{Security Practices}
	
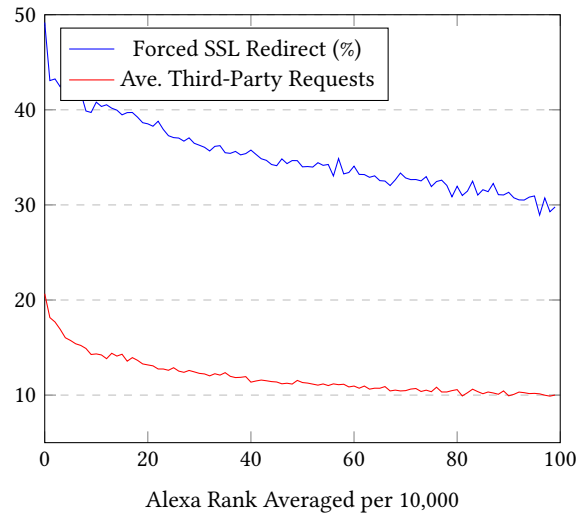
\begin{figure}
		\caption{Higher ranked sites force SSL more often (blue), but also initiate more third-party requests (red).}
\begin{tikzpicture}

	\begin{axis}[
		xlabel={Alexa Rank Averaged per 10,000},
		xmin=0, xmax=100,
		ymin=5, ymax=50,
		xtick={0,20,40,60,80,100},
		ytick=,
		legend pos=north west,
		ymajorgrids=true,
		grid style=dashed,
	]
 
\addplot[
    color=blue,
    mark=none,
    ]
    coordinates {
		(0,49.17721518987342)
		(1,43.075778509937294)
		(2,43.24584895554365)
		(3,42.426841033336906)
		(4,43.39704135622503)
		(5,42.59479312952361)
		(6,41.35306100568243)
		(7,41.7876975651431)
		(8,39.87598888176181)
		(9,39.721179624664884)
		(10,40.80683538827601)
		(11,40.352574438353216)
		(12,40.530099796115465)
		(13,40.17167381974249)
		(14,39.96576807873342)
		(15,39.47114869928273)
		(16,39.70572441198582)
		(17,39.71887977851134)
		(18,39.230438521066205)
		(19,38.64661654135338)
		(20,38.52502947797192)
		(21,38.27279545698061)
		(22,38.79763821792807)
		(23,37.936218189627404)
		(24,37.27886780315214)
		(25,37.07684040838259)
		(26,37.0295124037639)
		(27,36.704923307948086)
		(28,37.0521347350354)
		(29,36.481700118063756)
		(30,36.27608346709471)
		(31,36.05856577963023)
		(32,35.67226441330624)
		(33,36.15728900255755)
		(34,36.23017573939134)
		(35,35.487661574618095)
		(36,35.42202227934876)
		(37,35.625800939769334)
		(38,35.261472785485594)
		(39,35.39529914529915)
		(40,35.76493339063171)
		(41,35.3054045410937)
		(42,34.85592315901814)
		(43,34.6853595890411)
		(44,34.24745581146224)
		(45,34.125546200575506)
		(46,34.84051122328429)
		(47,34.34429730884238)
		(48,34.66254398123467)
		(49,34.65378079864061)
		(50,34.00342172797263)
		(51,34.03437600085406)
		(52,33.981510997768574)
		(53,34.438286079727135)
		(54,34.16168941979522)
		(55,34.251675353685776)
		(56,33.04227451815568)
		(57,34.8739946380697)
		(58,33.262100651779036)
		(59,33.40785691472373)
		(60,34.073678590496534)
		(61,33.20847602739726)
		(62,33.19117176671287)
		(63,32.90260572404955)
		(64,33.05732484076433)
		(65,32.553668695930796)
		(66,32.509891990161485)
		(67,32.03666204838538)
		(68,32.63303828516583)
		(69,33.35123523093448)
		(70,32.837910608508345)
		(71,32.660040611307046)
		(72,32.669751425804364)
		(73,32.52702558064862)
		(74,32.97541071620316)
		(75,31.92731992212849)
		(76,32.458697764820215)
		(77,32.60351181443746)
		(78,32.05058912549995)
		(79,30.856832971800436)
		(80,31.960443381873503)
		(81,30.990833697075516)
		(82,31.449928547872926)
		(83,32.50027403266469)
		(84,31.03448275862069)
		(85,31.597606913361396)
		(86,31.393925909444874)
		(87,32.25225225225225)
		(88,31.07709750566893)
		(89,31.046054154130985)
		(90,31.324029553184)
		(91,30.748120749313923)
		(92,30.52887537993921)
		(93,30.51016589746788)
		(94,30.82077051926298)
		(95,30.941824533118368)
		(96,28.934449898931565)
		(97,30.731936190178295)
		(98,29.28205128205128)
		(99,29.783754315827732)
    };
   	\legend{Forced SSL Redirect (\%), Ave. Third-Party Requests}
 \addplot[
    color=red,
    mark=none,
    ]
    coordinates {
		(0,20.65763406580895)
		(1,18.151277013752456)
		(2,17.686686802973977)
		(3,16.92178447276941)
		(4,16.032205176250148)
		(5,15.746830985915492)
		(6,15.398535564853557)
		(7,15.221012394300939)
		(8,14.901708401591387)
		(9,14.270984334814122)
		(10,14.331640486595015)
		(11,14.223261250730568)
		(12,13.830144657022865)
		(13,14.396881228748974)
		(14,14.10344426343961)
		(15,14.296451425247236)
		(16,13.566997084548104)
		(17,13.94197359277276)
		(18,13.656928311627361)
		(19,13.279475471256292)
		(20,13.174352421884075)
		(21,13.076743915220682)
		(22,12.746829497416627)
		(23,12.744633431085044)
		(24,12.610243157523787)
		(25,12.880221202494411)
		(26,12.522551998130405)
		(27,12.395470993346562)
		(28,12.591879840413048)
		(29,12.443518083607328)
		(30,12.285330216247809)
		(31,12.225219941348973)
		(32,12.023681754549697)
		(33,12.245579339227548)
		(34,12.103452325035228)
		(35,12.359109548916228)
		(36,11.9734492481203)
		(37,11.842837111474644)
		(38,11.869290231904428)
		(39,11.941761197520758)
		(40,11.361724748966331)
		(41,11.48694532256176)
		(42,11.581996956572633)
		(43,11.508695652173913)
		(44,11.419434628975266)
		(45,11.378296382730456)
		(46,11.189122518501115)
		(47,11.247997172478794)
		(48,11.160892542571933)
		(49,11.544449616011171)
		(50,11.320216776625825)
		(51,11.262660443407235)
		(52,11.154909936083673)
		(53,11.046517036864751)
		(54,11.173227426702487)
		(55,11.003028890959925)
		(56,11.180992313067785)
		(57,11.0958564514225)
		(58,11.140562719812428)
		(59,10.859205776173285)
		(60,10.951228029333024)
		(61,10.725076596747584)
		(62,10.95270588235294)
		(63,10.62533756017377)
		(64,10.726126546812981)
		(65,10.720216394213807)
		(66,10.896734645762113)
		(67,10.43813834726091)
		(68,10.524004205116224)
		(69,10.438309394866964)
		(70,10.469344359125811)
		(71,10.633356873822976)
		(72,10.692997992679183)
		(73,10.380986081623025)
		(74,10.518860115880337)
		(75,10.35693950177936)
		(76,10.826459051467966)
		(77,10.32770511296076)
		(78,10.328873491816987)
		(79,10.471452460582896)
		(80,10.577429392053615)
		(81,9.91115107913669)
		(82,10.246950851346456)
		(83,10.623015873015873)
		(84,10.365037001091835)
		(85,10.145031660983927)
		(86,10.326086956521738)
		(87,10.226476072094469)
		(88,10.087879166146548)
		(89,10.441108840061318)
		(90,9.923891975707456)
		(91,10.077659714023351)
		(92,10.318151373074347)
		(93,10.255096418732782)
		(94,10.157820039738858)
		(95,10.18175085172567)
		(96,10.13276299112801)
		(97,9.990187639869168)
		(98,9.89020715630885)
		(99,10.00624496373892)
    };
\end{axis}
\end{tikzpicture}
\end{figure}
	
	In addition to notice and choice, guidelines for online advertising often include provisions for ensuring data security.  U.S. Federal Trade Commission online advertising guidelines from 2009 assert that ``Any company that collects and/or stores consumer data for behavioral advertising should provide reasonable security for that data''\cite{ftc-2009-oba}.  Likewise, the ``Self-Regulatory Principles for Online Behavioral Advertising'' authored by industry trade group Internet Advertising Bureau dictate that: ``Entities should maintain appropriate physical, electronic, and administrative safeguards to protect the data collected and used for Online Behavioral Advertising purposes'' \cite{iab-2010-self_reg}.

	In the context of third-party data transfer on the web, there are two main technical factors involved in protecting user data: storage encryption and transport encryption.  Storage encryption applies to how data is protected once it reaches its destination and protects against unauthorized parties reading data after it has been received and processed.  Transport encryption refers to the process by which data is encrypted as it is transferred over the Internet and protects against network adversaries reading the data.
	
	It is impossible to verify if third-parties collecting user data are employing sufficient storage encryption without an independent auditing body.  At present, no such body provides publicly-available reports of security practices.  However, it is possible to determine if transport encryption is being used by examining the network traffic generated when loading a page in order to determine if connections are made utilizing Secure Sockets Layer (SSL) connections.
	
	While transport encryption adoption has been increasing, there is still a large volume of unencrypted traffic which places user data at risk of interception.  Of all pages examined, 35.14\% redirect users to an SSL-secured HTTPS page after being requested via an HTTP request.  As Figure 3 illustrates, higher-ranked websites force SSL connections more often.  In terms of page content, 52.25\% of all element requests are encrypted.  However, there is significant variability in encryption between first- and third-party requests: first-party are encrypted in 35.52\% of cases compared to 66.82\% for third-parties.

	The above findings suggests that third-parties may have superior data security practices.  However, as Table 1 illustrates, there is huge variability in the encryption practices of the 25 examined third-party data collectors.  Facebook, Twitter, and New Relic all encrypt over 90\% of requests.  In contrast, Oracle, Acxiom, Lotame, Neustar, Nielsen Online, and Quantcast encrypt fewer than half of all requests.  This wide variability underscores how self-regulation produces wildly differing practices among data collectors and suggests that clear standards should be adopted and enforced.

\section{Prior Work}

	This study builds directly on three established lines of research, those investigating web privacy and third-party tracking, those addressing privacy policy usability, and those addressing legal and normative objections to notice and choice.

	Web privacy and third-party tracking is an active area of research.  Krishnamurthy and Wills conducted several early censuses of third-party tracking \cite{krishnamurthy-2006-generating}.  More recently both Libert and Englehardt and Narayanan have investigated tracking on the Alexa one million most popular sites \cite{libert-2015-1m-tracking,englehardt-2016-million_track}.  Several studies have looked at the longitudinal evolution of third-party tracking on the web \cite{krishnamurthy-2009-privacy,lerner-2016-internet}, defenses against tracking \cite{roesner-2012-detecting}, the economics of online tracking \cite{gill-2013-best,budak-2016-understanding}, tracking on smartphone apps \cite{van-2017-better,zang-2015-knows}, and the prevalence of different tracking methods \cite{acar-2014-web,bashir-2016-tracing,cahn-2016-empirical}.  Beyond general censuses, many studies have looked at specific implications of third-party tracking: Libert has investigated tracking on health-related webpages \cite{libert-2015-healthtracking}, Englehardt et al have investigated the nexus between web tracking and state-sponsored surveillance \cite{englehardt-2015-cookies}, and Hauschke has detailed tracking on library websites \cite{hauschke-2016-third}.  

	Numerous studies have investigated usability aspects of the notice and choice policy framework.  McDonald and Cranor investigated the time required to read policies and the overall cost to national productivity \cite{mcdonald-2008-cost}.  Likewise, McDonald et al have conducted comparative work on privacy policies \cite{mcdonald-2009-comparative}.  Another study from Cranor et al ``automatically evaluated 6,191 U.S. financial institutions' privacy notices'' \cite{cranor-2016-large}.  Reidenberg et al investigated the degree to which users have trouble understanding policies \cite{reidenberg-2014-disagreeable}, while Acquisti and Grossklags have found that ``even if individuals had access to complete information...they might still deviate from the rational strategy'' \cite{acquisti-2005-privacy}.   Wilson et al have evaluated the feasilibty of crowdsourcing annotation of privacy policies \cite{wilson-2016-crowdsourcing}.  Komanduri et al found that members of online advertising industry groups the Network Advertising Initiative (NAI) and Digital Advertising Alliance (DAA) did not always follow their own notice and choice guidelines \cite{komanduri-2011-adchoices}.

	Legal and normative scholars have extensively addressed fundamental issues with the notice and choice policy regime.  Solove has stated that the regime ``cannot achieve the goals demanded of it, and it has been pushed beyond its limits'' \cite{solove-2012-privacyselfmanagement}.  Cate has criticized the fact that the FTC has chosen not to regulate privacy practices, but instead ``focused virtually all of its...efforts on getting websites to post privacy policies and its enforcement efforts on suing website operators when they fail to follow those policies'' \cite{cate-2006-consumer}.  Barocas and Nissenbuam have observed that ``users who are subject to [online tracking] confront not only significant hurdles but full-on barriers to achieving meaningful understanding of the practice and uses to which they are expected to be able to consent'' \cite{barocas-2009-notice}.  Rotenberg has asserted that the notice and choice paradigm is ``a relatively recent creation of the U.S. marketing industry'' and is at odds with internationally recognized data protection frameworks \cite{rotenberg-2001-fair}.

\section{Conclusion}

	This study has shown empirically for the first time that the notice and choice policy regime fails to notify users reading privacy policies of the parties which collect their data.  It also demonstrates that the time burden for reading \emph{both} first- and third-party policies is unmanageable.  Furthermore, the ``Do Not Track'' mechanism is rarely mentioned in privacy policies, and when it is, it is usually to specify the expressed user choice is ignored.  Finally, while implementing SSL encryption is an easy means of ensuring transport security, there is uneven support across the third-parties which collect user data on the web.

	Private-sector technologists routinely assert that privacy is a modern invention which may be viewed as an ``anomaly'' \cite{verge-2013-cerf_privacy}.  Such a viewpoint is not only ignorant of widely documented instances where privacy has been endorsed as a foundational social value by numerous ancient cultures \cite{nih-2002-hippocratic,acquisti-2015-privacy}, it is also ignorant of the \emph{recent} history of data protection regulations around the world \cite{gellman-2016-fair}.  Furthermore, decades of survey research have consistently demonstrated that online privacy is valued by the public \cite{cranor-2000-beyond,gandy-2003-public,turow-2007-internet,pew-2014-privacy_post_snowden,turow-2015-tradeoff_fallacy}.  Thus, the true anomaly may be that a massive sector of the global economy has been regulated by a fundamentally broken approach which is sharply at odds with public desires.

\section{Acknowledgements}

	I would like to thank the reviewers for their helpful comments, Reuben Binns and Nataliia Bielova for assistance with the abstract, and my dissertation committee who oversaw earlier versions of this work (Victor Pickard, Jonathan M. Smith, Michael X. Delli Carpini, Joe Turow, Guobin Yang).  Jonathan M. Smith deserves additional thanks for providing the computational resources needed for this project well beyond my disseration defense.

\bibliographystyle{ACM-Reference-Format}
\balance
\bibliography{sample-bibliography}

\end{document}